\documentclass[ prd,aps,showpacs,nofootinbib,floatfix,twocolumn ]{revtex4-1}
\usepackage{slashed}
\usepackage{float}
 \usepackage{amsmath,graphicx,color,epsfig,ulem}
\usepackage{amssymb}
 \usepackage{CJK}
\begin{document}

\title{Nambu--Jona-Lasinio model with proper time regularization in a finite volume }
\author{Qing-Wu Wang $^{1}$}~\email[]{Email: qw.wang@scu.edu.cn}
\author{Yonghui Xia$^{2}$}
\author{Hong-Shi Zong$^{2,3,4}$}~\email[]{Email: zonghs@nju.edu.cn}

\affiliation{
$^1$Department of Physics, Sichuan University,  Chengdu 610064, China\\
$^2$ Department of Physics, Nanjing University, Nanjing 210093,
China\\
$^3$ Joint Center for Particle, Nuclear Physics and Cosmology, Nanjing 210093, China \\
$^4$ State Key Laboratory of Theoretical Physics, Institute of Theoretical
Physics, CAS, Beijing 100190, China
 }

\begin{abstract}
 Based on the two flavor NJL model with a proper time regularization, we used stationary wave condition (SWC) for the first time to study the influence of the finite volume effects on the chiral phase transition of quark matter at finite temperature. It is found that when the cubic volume size $L$ is large than $L_{max}^{SWC}=500$ fm, 
the chiral quark condensate is indistinguishable from that at $L=\infty$. Here it should be noted that $500$ fm is far greater than the size of QGP produced at laboratory and the lattice QCD simulation space size. It is also much larger than the previous limit size $L_{max}^{APBC}=5$ fm estimated by the commonly used anti-periodic boundary condition (APBC). We also found that when the space size $L$ is less than $L_{min}^{SWC}=0.25$ fm, the spontaneous symmetry breaking concept is no longer valid. In addition, we first introduce the spatial susceptibility, and through the study of the spatial susceptibility, it was revealed for the first time that the chiral phase transition caused by the finite volume effects in the non-chiral limit is a crossover.

\end{abstract}

\maketitle

\section{Introduction}
 The quark-gluon plasma (QGP)
is widely believed as a critical state in the early universe may now be reproduced through relativistic heavy ion collisions (RHICs) \cite{Adams:2005dq,Shuryak:2008eq}.
However, it should be noted that the QGP system produced at RHICs always has a finite size. For example, the authors of Refs. \cite{Bass:1998qm,Graef:2012sh} pointed out that the volume of homogeneity before freeze out for $Au-Au$ and $Pb-Pb$ collisions range between approximately $50 \sim 250$ fm$^3$, and the authors of Ref. \cite{Palhares:2009tf} estimated that the volume of the smallest QGP system could be as low as $(2 fm)^3$.
Theoretically, in  QGP color deconfinement and chiral symmetry restoration have been identified \cite{Bilgici:2008qy,Fischer:2009jm,Endrodi:2011gv}  and finite volume effects are investigated through different methods including chiral perturbation theory \cite{Gasser:1986vb, Hansen:1990, Damgaard:2009}, Dyson-Schwinger approach \cite{Luecker:2009bs,Li:2017zny,Shi:2018chao,Shi:2018SC}, Polyakov-loop extended Nambu$-$Jona-Lasinio model \cite{Bhattacharyya:2015,Bhattacharyya:2013,Pan:2017},  quark-meson model \cite{Braun:2004yk,Braun:2005fj,Colangelo:2005gd,Colangelo:2010ba}  and other non-perturbative renormalization group method.  It is found that the finite volume affects the critical behavior at the chiral phase transition of quark matter. The latest summary paper on the study of finite volume effects can be found in Ref. \cite{Klein:2017shl}.

As mentioned above, the scale of the QGP produced at RHICs is limited. In order to study finite volume effects on the critical behavior of quark matter, one need to determine the boundary conditions of the QGP produced at RHICs in advance. Theoretical studies show that the chiral condensate and meson mass  rely on the choice of boundary conditions
\cite{Braun:2005gy,Carpenter:1984dd,Fukugita:1989yw,Aoki:1993gi}.
In the virtual time temperature field theory, temperature is introduced by replacing integration along the temporal direction with a summation of the Matsubara frequency which satisfies the anti-periodic boundary conditions for a fermion field. But, unlike the thermal Matsubara frequencies, which are fixed by the statistics of the fields, there is no such restriction for the boundary conditions in the spatial directions. So an important question came up, for a limited size physical system, how to select the boundary conditions of the fermion fields in the spatial direction?

For a limited size physical system, the prevailing boundary conditions in QCD include the  periodic boundary condition (PBC) and anti-periodic boundary condition (APBC) (Of course, there are some other boundary conditions, such as a MIT boundary condition \cite{Chernodub:2016kxh,Chernodub:2017mvp,Chernodub:2017ref}). In Refs. \cite{Klein:2017shl,Gasser:1987ah,Jungnickel:1997yu}, it claims that in spatial directions the fields should  take  the same boundary condition as that in the temporal direction for a practical effective QCD model.
So anti-periodic boundary is the first choice for many effective QCD model studies. But most current lattice QCD simulations still favor a periodic quark boundary condition as a \textit{ de facto} standard \cite{Aoki:1993gi,Klein:2017shl}. In this case, here a natural problem appears: why does the same finite size system adopt different boundary conditions for different methods? Are the commonly used conditions of APBC or PBC suitable for studying finite size systems?


A fireball (plasma) created in laboratory experiment is restricted to a finite (small) volume surrounded by a cold exterior.
This means that quark fields in the hot plasma are not likely to escape the cold exterior, that is quark`s wave function must be equal to zero on the boundary. Evidently, the commonly used conditions of APBC or PBC in the past can not satisfy this requirement. In that case we first put forward the application of stationary wave condition (SWC) to study the finite-volume effects of QCD chiral phase transition, this is because that only the SWC can satisfy the requirement that  quark`s wave function is equal to zero on the boundary \cite{Wang:2018ovx}. Here, it should be stressed that SWC is quite different from APBC or PBC for the studying of finite size system, this will be clearly reflected in the following calculations.

Once the boundary conditions are selected, the selected boundary conditions can be used to study the finite volume effects on the chiral phase transition of QCD. Before we do this, it is very useful to review the conditions for the establishment of the chiral symmetry spontaneous breaking. It is well known that spontaneous symmetry breaking actually occurs only for idealized systems that are infinitely large, which means that finite volume system does not exist the phenomenon of  spontaneous symmetry breaking in principle. But all real physical systems are limited, in this case, we want to ask how large a physical system ($L_{max}$) can regard as an ideal infinite thermodynamic system? Furthermore, we want to know how small the physical systems ($L_{min}$) is the symmetry spontaneous breaking is no longer present? The main motivation of this paper is to try to answer the above questions under the framework of Nambu--Jona-Lasinio model (NJL). In addition, in this paper we further discuss the relationship between time discretization and spatial discretization in Euclidean space.

For non-renormalizable effective model, such as NJL model, a procedure for regulating divergent quantities is required.  In fact, there are a lot of regularization schemes, each regularization scheme has its strengths and weaknesses.  The commonly used NJL model generally uses  three-momentum ultraviolet (UV)-cutoff of the  momentum space to regularize the amount of the divergence. However, it should be pointed that this 
regularization scheme is not suitable for studying finite volume effects. The specific reasons are as follows: 
It is well known that in the regularization scheme of three-momentum cutoff $\Lambda$ is generally selected to be 600 $MeV$. In order to study the small volume effects, the momentum space needs to be discretized. Once the momentum space is discretized, it means that the contribution of high-frequency mode will be ignored due to UV-cutoff $\Lambda$, which is particularly disadvantageous for the study of the small-volume effects. In this case, we have to abandon the common used three-momentum cutoff and instead use the  proper-time regularization to study the finite size effects. This is because the  proper-time regularization is not plagued by the interruption of UV momentum. The proper-time regularization 
was first proposed by J. Schwinger \cite{Schwinger} and widely used to study the properties of hadron (see Ref. \cite{Vogl} and therein) and chiral phase transition \cite{Menezes,Ayala,Ferrera}. This regularization scheme shows an obvious advantage: all the symmetry is kept.

In Refs. \cite{Tripolt,Juri}, it has found that chiral behaviors are depended on the UV-cutoff and meson mass in the quark-meson model. As work with the (anti-) period boundary condition,  a convenient way is to  use the proper time regularization \cite{Liao:1994fp,Litim:2001hk,Zappala:2002nx,Cui:2014hya,Zhang:2016zto}.
Therefore, in this work, we will adopt the  proper-time regularization instead of the usual three-momentum cutoff to study the small volume effects on the chiral phase transition.

The paper is organized as follows. In Sec. \ref{sec.model}, we will introduce the NJL model with proper time regularization and give the corresponding model parameters. In Sec. \ref{sec.results}, we give
two different calculation results under the condition of APBC and SBC and further analyze the equivalence relationship between  $T$ and $1/L$.
Finally, a summary is given in Sec. \ref{sec.summary}.

\section{NJL Model with proper time regularization }
\label{sec.model}


The deconfined quark matter within NJL model with four-quarks interaction in Euclidean space can be described by  Lagrangian \cite{Klevansky:1992qe,Buballa:2003qv}
\begin{eqnarray}
\mathcal{L}_{NJL} & =&\bar \psi (i\gamma_\mu)\partial^\mu-\hat m_q\psi  \nonumber\\
 &+&G[(\bar \psi  \psi)^2+(\bar \psi i\gamma_5\tau  \psi)^2],
\end{eqnarray}
where $G$ is the   effective coupling and $\hat m_q$ is the mass matrix.  We   consider the two flavors u and d quarks   with exact isospin symmetry.

 In the mean field approximation, the dressed quark mass $M$ is defined through the chiral condensate with  $
 M=m+\sigma$ in which
 \begin{equation}
 \sigma=-2G\left\langle {  \bar \psi \psi} \right\rangle
 \end{equation}
 and    the  condensate is defined as
\begin{equation}\label{eq.gap1}
\left\langle {  \bar \psi \psi} \right\rangle=-\int \frac{d^4p}{(2\pi)^4}Tr[S(p)].
\end{equation}
Here $\psi=(\psi_u,\psi_d)^T$ with the number of color $N_c=3$ and the number of flavor $N_c=2$. $S(p)$ is the dressed quark propagator and the trace is taken in color, flavor and Dirac spaces.

In the framework of the proper time regularization, for a free particle propagator with mass $M$, we have
  \begin{eqnarray}
  \frac{1}{p^2+M^2}&= &  \int_{0}^ \infty d\tau  e^{-\tau (p^2+M^2)} \nonumber\\
  & \rightarrow & \int_{\tau_{UV}}^ {\infty} d\tau  e^{-\tau (p^2+M^2)}.
   \end{eqnarray}
Here $\tau_{UV}=1/\Lambda_{UV}^2$ is the ultraviolet (UV) cutoff for the proper time regularization.

With the proper time regularization, the  quark condensate Eq. (\ref{eq.gap1}) in the infinite-volume limit and  at zero temperature  can be written as
\begin{eqnarray}\label{eq.condensate}
\left\langle {  \bar \psi \psi} \right\rangle&=&-N_cN_f\int\frac{d^4p}{(2\pi)^4}\frac{4M}{p^2+M^2}  \nonumber \\
&=&-24M\int^\infty_{-\infty} \frac{d^4p}{(2\pi)^4}\int^ \infty_{\tau_{UV}} d\tau e^{-\tau(p^2+M^2)}  \nonumber\\
&=&-\frac{3M}{2\pi^2}\int^ \infty_{\tau_{UV}} d\tau\frac{ e^{-\tau M^2}}{\tau^2}.
\end{eqnarray}

 As for finite temperature, the quark four-momentum is  replaced by $p_k=(\vec{ p}, \omega_k)$, with $\omega_k=(2k+1)\pi T$, $k \in  \mathbb{Z}$. 
The fourth momentum is replaced by a sum of all the fermion Matsubara frequencies $\omega_k$. Explicitly, it is a replacement
 \begin{equation}\label{eq.dT}
 \int  dp_4 e^{-\tau p_4^2}\rightarrow 2\pi T\sum  _{k}e^{-4\tau (k+\frac{1}{2})^2},\quad k= \pm 1,\pm 2...
 \end{equation}
In the case of finite temperature, the temperature-dependent two-quark condensate can be written as
\begin{eqnarray}\label{eq.cons2}
\left\langle {  \bar \psi \psi} \right\rangle&=&-T\sum^\infty_{k=-\infty}\int \frac{d^3p}{(2\pi)^3}Tr[G(p_k^2)]  \nonumber\\
\rightarrow&-&24M\int^ \infty_{\tau_{UV}} d\tau e^{-\tau M^2}[T\sum^\infty_{k=-\infty}\int^\infty_{0} \frac{dp}{2\pi^2}p^2e^{-\tau( p^2+\omega_k^2)}]   \nonumber  \\
&=&-\frac{3MT}{\pi^{3/2}}\int^  \infty_{\tau_{UV}} d\tau\frac{ e^{-\tau M^2}}{\tau^{3/2}}\theta_2(0,e^{-4\pi^2\tau T^2}),
\end{eqnarray}
where the Jacobi function is defined as $\theta_2(0,q)=2\sqrt[4]{q}\sum\nolimits_{n=0}^\infty q^{n(n+1)}.$
Then the dressed quark mass  is
\begin{eqnarray}\label{gapint}
M=m+\frac{6GMT}{\pi^{3/2}}\int^  \infty_{\tau_{UV}} d\tau\frac{ e^{-\tau M^2}}{\tau^{3/2}}\theta_2(0,e^{-4\pi^2\tau T^2}).
\end{eqnarray}
Here the commonly used parameters for the proper-time regularization  are $m=5$ MeV, $G=3.26*10^{-6}$ MeV$^{-2}$, and $ \Lambda_{UV}=1080$ MeV which give  the dressed quark mass $ M=223.7$ MeV at zero temperature. As has been shown in Ref. \cite {Cui:2013aba}, under the framework of the proper-time regulation, all model calculations results are insensitive to the model parameters used. In addition, we assume in this paper that the coupling $G$ is a constant that does not depend on the volume size $L$ and other physical quantities \cite{Wang:2016fzr,Cui:2014hya,Ayala:2016bbi,Li:2018}.

\section{Results}
\label{sec.results}
\subsection{Anti-period boundary condition}

In a limited size physical system, if the quark momentum is  discretized with APBC, then the momentum is replaced by
\begin{eqnarray}\label{eq.apbc}
\vec{ p} ^2&=&\frac{4\pi^2}{L^2}\sum\nolimits_{i=1}^{3}(n_i+\frac{1}{2})^2,\quad n_i= \pm 1,\pm 2...
 \end{eqnarray}
 while the integral measure is replaced by sum of  discrete momentum with
\begin{equation}\label{eq.dp}
\int  dp_i e^{-\tau p_i^2}\rightarrow\frac{2\pi}{L}\sum_{ n_i=-\infty  }^{\infty}e^{-\tau p_i^2}. \qquad i=1,2,3.
\end{equation}
Here the sum with zero mode will be replaced by a $\theta_2$ function as in Eq. (\ref{eq.cons2}).

Comparing Eq. (\ref{eq.dT}) with  Eqs. (\ref{eq.apbc},\ref{eq.dp}), we can easily find that one-dimensional time and any one-dimensional momentum space in 
three-dimensional momentum space are discretized in the same way, that is, one-dimensional time and any one-dimensional space in three-dimensional space here are discretized with APBC, we will get the following  mathematical equivalence for a finite physical system 
\begin{equation}\label{eq.TL}
T \sim \frac{1}{L}.
\end{equation}
This is not surprising, due to the fact that in this case, the Euclidean space and time directions are physically equivalent, and a permutation symmetry holds among them.
However, it should be noted here that if the time dimension and the spatial dimension are discretized in different ways, for example, the spatial dimension is discretized according to SWC, then the equivalence relationship between $T \sim \frac{1}{L}$ will have interesting changes. See the next section for details.

It is obvious that the temperature effect can be factor out from the momentum integration as in Eq. (\ref{gapint}). If the spatial and temporal directions are both taking the anti-boundary condition, the sums of the discretized variables   give the same $\theta_2$ function while the integral in continuum space direction give a term $1/(2\sqrt{\pi\tau})$.
Then with a replacement of $T \to 1/L$, the chiral condensate Eq. (\ref{eq.cons2}) can be viewed as a result of zero temperature but discretized in one spatial direction which means the temperature effect may be replaced by finite volume effect.

The time dimension of Euclidean space can be discretized in only one-dimension, but we can arbitrarily discretize one-dimensional, two-dimensional, or three-dimensional momentum in three-dimensional momentum space.
In order to better reflect the influence of the limited space size in three-dimensional space, we will study the chiral behavior with  discretization of spatial directions in arbitrary number of dimensions.

For a two flavor quark system of finite volume  with $n$ continuum  and $k$ discretized directions, the dressed quark mass is constrained by
\begin{eqnarray}\label{eq.gapnk}
M&=&m+48GM\int^ \infty_{\tau_{UV}} d\tau e^{-\tau M^2}(\frac{1}{2\sqrt{\pi \tau}})^n\times  \nonumber \\
&&[T \theta_2(0,e^{-4\pi^2\tau T^2})]^\alpha[\frac{\theta_2(0,e^{-4\pi^2\tau/L^2})}{L}]^{(k-\alpha)}
\end{eqnarray}
with $\alpha=1 ($ for $ T>0)$, or $0 ($ for $ T=0)$ and $n+k=4$. It is easy to see from Eq. (\ref{eq.gapnk}) that the discretization of the time dimension and any one-dimensional momentum space in three-dimensional momentum space are completely equivalent to the dressed quark mass. What needs to be pointed out here is, it is because of the  proper-time regularization and APBC used in this paper that we can get the equivalence between $T$ and $\frac{1}{L}$ very directly.

\begin{figure}
{\includegraphics[width=0.8\columnwidth]{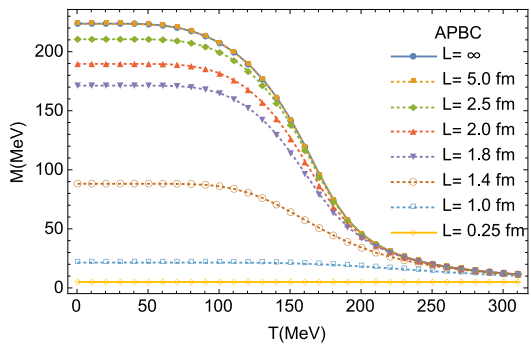} }
\caption{ The dressed quark mass as function of temperature for different volume size with anti-period boundary condition. The quark mass  decreases as $L$ decreases below $5$ fm.
 \label{fig.mapbc}}
\end{figure}

The chiral behavior is depended on the size and shape of the restraint system. It is important in extracting chiral transition information in heavy ion collisions as the ensembles may have differently shape and  size. In this work, we will study the case that the three spatial directions are not all finite.
If one-dimensional temporal and three-dimensional spatial directions are all discretized, but $L$ and $T$ are changed independently, it corresponds to the case  $(n, k)=(0, 4)$.
 In Fig. \ref{fig.mapbc}, we drew the variation of the dressed quark mass with temperature in different space size. Just as shown in Fig. \ref{fig.mapbc}, the dressed quark mass decreases with decreasing space size and the dynamical chiral symmetry is total restored in very small volume. Fig. \ref{fig.mapbc} clearly shows that when the cubic volume size $L$ is large than $L_{max}^{APBC}=5$ fm, the limited physical system reaches the infinite volume limit (when $L\geq L_{max}^{APBC}$, the chiral quark condensate is indistinguishable from that at $L=\infty$).
This means that when the space size $L$ is greater than $L_{max}^{APBC}=5$ fm, the corresponding finite size physical system can be regarded as an ideal infinite thermodynamical system. Here, it is  beneficial to compare our results by means of the NJL model with proper-time regularization to that of other non-perturbative methods, such as Dyson-Schwinger approach . 
In our work, $L^{APBC}_{max}=5$ fm given by APBC, and in the Dyson-Schwinger approach, $L^{APBC}_{max}=3$ fm \cite{Shi:2018chao}, comparing these two results can be found quantitatively comparable to each other.
Fig. \ref{fig.mapbc} also tell us when the space size is less than $L_{min}^{APBC}=0.25$ fm, the spontaneous symmetry breaking concept is no longer valid, this is because the dressed quark mass and the corresponding current quark mass completely equal in this case. In view of the size of QGP produced at RICHs between  $50 \sim 250$ fm$^3$ \cite{Bass:1998qm,Graef:2012sh,Palhares:2009tf}, the results given by APBC show that the finite volume effects can not be ignored.

Now let's discuss more concretely the relationship between time discretization and spatial discretization in Euclidean space. If we replace the $T$ with $1/L$, the gap equation Eq. (\ref{eq.gapnk}) can be viewed in two ways:
  \begin{itemize}
  \item{$T=0$,  with $k$ spatial dimensions are discretized; }
  \item{$T=1/L$, with $k-1$ spatial dimensions are discretized.}
  \end{itemize}

Then the case $(n,k)_{\alpha=0,T=0}$ is equal to the case $(n,k)_{\alpha=1,T=1/L}$. For example, we consider the following scenario:
 \begin{equation}
\sum_{T=0}\sum_{L_x,L_y\to \infty   }\sum_{L_z \to  \text{ finite}} \sim \sum_{T>0}\sum_{L_x, L_y, L_z\to \infty},
\end{equation}
where $L_i$ is the size of $i'{th}$ spatial direction.
Our numerical calculation shows that the chiral phase transition occurs at $L=1.2$ fm or at pseudo-critical temperature  $T=165$ MeV which equal to the Dyson-Schwinger approach value and  close to the lattice simulation value $T_c=154~ (9)$ MeV \cite{Bazavov:2012}. So, in this sense we can say that the effect of temperature on chiral phase transition is completely equivalent to the effect of the finite space size (one dimension).

\begin{figure}[H]
{\includegraphics[width=0.8\columnwidth]{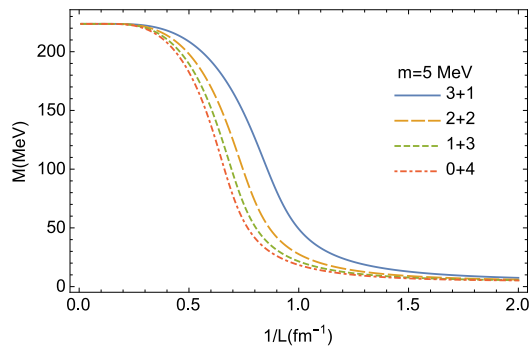} }
\caption{ Quark mass as function of  $1/L$.
The notation $(n+k)$ with $n+k=4$ means   $k$ directions are discretized.
 }\label{fig.m123}
\end{figure}

\begin{figure}[H]
{\includegraphics[width=0.8\columnwidth]{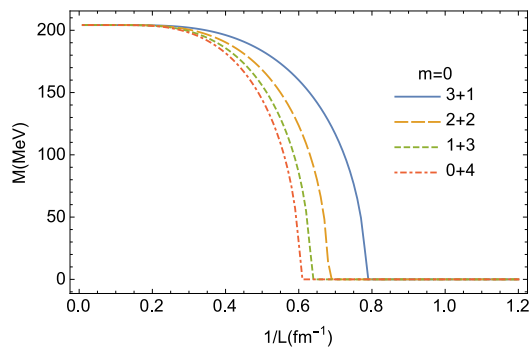} }
\caption{Quark mass as function of $1/L$ in the chiral limit. The transition point are at $L=1.64$, $1.56$, $1.45$, $1.27$ fm for $k=4$, $3, 2, 1$ respectively.
 }\label{fig.chim0}
\end{figure}

The results for the dressed quark mass  in different $n$ and $k$ are showed in Fig. \ref{fig.m123}. From
Fig. \ref{fig.m123}, it is easy to find that the more discretized in the Euclid space dimensions is, the faster the dressed quark mass declines. The  temperature effect is not naively equivalent  with volume effect.
Similarly we can discuss the chiral limit. In the chiral limit, we fix the effective coupling and ultraviolet cutoff unchanged which give a dressed quark mass of $204$ MeV at zero temperature. The corresponding results are  showed in Fig. \ref{fig.chim0}. The dressed quark mass varies with the volume reveals a second order phase transition in the chiral limit. The chiral phase transition at the chiral limit induced by the finite volume effects is very similar to the phase transition caused by the finite temperature effect.

\begin{figure}[H]
{\includegraphics[width=0.8\columnwidth]{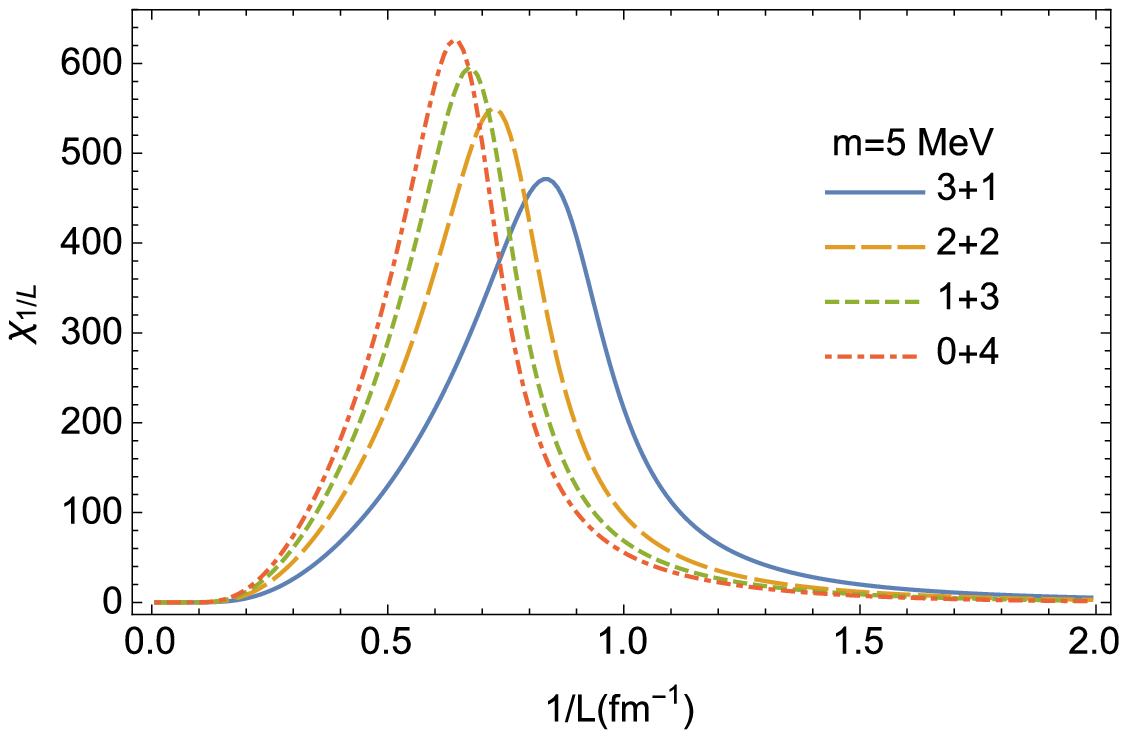}\\
\includegraphics[width=0.8\columnwidth]{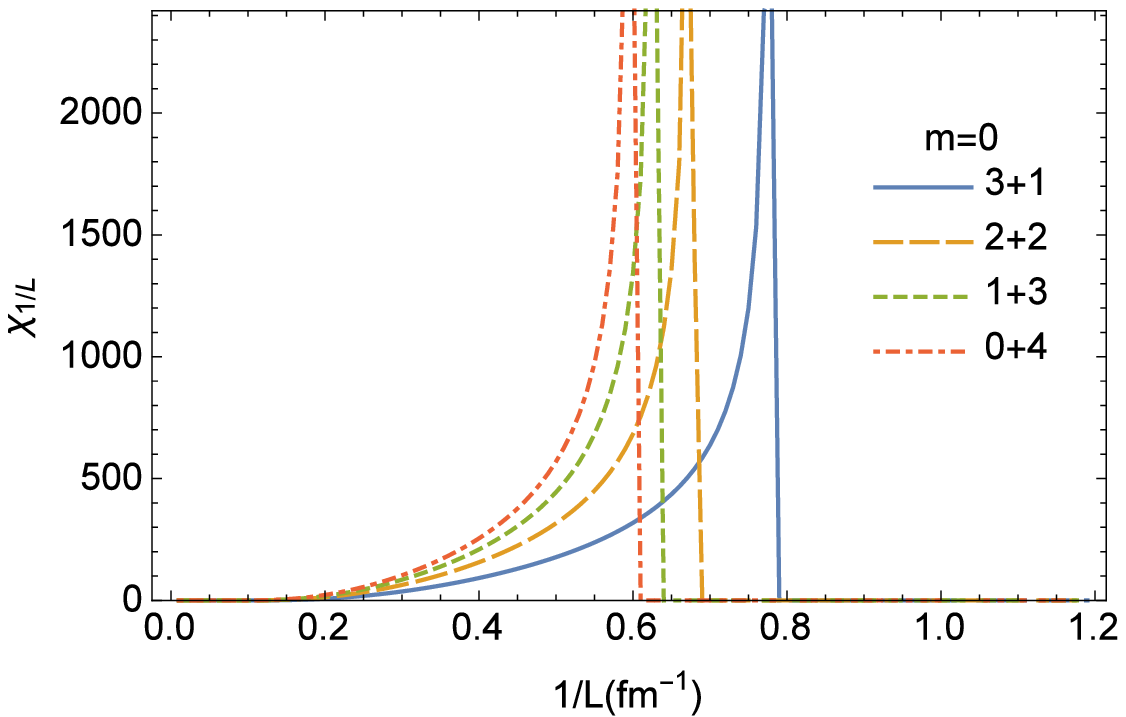} }
\caption{ The spatial susceptibilities as function of $1/L$. }
\label{fig.x123}
\end{figure}

In order to quantitatively reflect the finite volume effects on the QCD chiral phase transition, we  have derivative of the  chiral quark condensation with respect to spatial size $1/L$. It is a new vacuum susceptibility which similar to the susceptibility for the  chiral quark condensation with respect to temperature. We call it spatial susceptibility which reads as
\begin{equation}
\chi_{1/L}(L)=-\frac{\partial \sigma}{\partial (1/L)}.
\end{equation}
The results of spatial susceptibility with different $(n,k)$ are showed in Fig. \ref{fig.x123}.

From Fig. \ref{fig.x123}, it can be clearly seen that the chiral phase transition caused by the small volume effect is a second-order phase transition and a crossover in the case of chiral limit and non-chiral limit, respectively. This phenomenon is very similar to the chiral phase transition caused by the finite temperature. What needs to be pointed out here is that no one has discussed the characteristics of the chiral phase transition due to the small-space size effect in the previous studies, and this is where the meaning of the spatial susceptibility.

The finite volume effect can be presented in two ways. Firstly,
 comparing the Fig. \ref{fig.m123} with Fig. \ref{fig.x123},
for a given volume size $L$, we can see that the more the space-time dimension is discretized, the smaller the dressed quark mass becomes, and at the same time the position of the chiral transition point $L$ increases.
 With the reduction of volume size $L$, the dressed quark mass $M$ decreases, and dynamical chiral symmetry is partially restored.
Secondly, the plots for $k\geq2$ ($T \neq 0$ and at least one spatial direction is discretized ) can be viewed as
temperature   increases as the volume decreases with $T=1/L$.

The phase transition size $L$ and corresponding temperature are listed in  Tab. {\ref{tab.tab1}}.
 \begin{table}[h]
 \caption{The chiral phase transition position $L$ and the corresponding temperature  in the four cases of  discretized dimensions. The meaning of $n$ and $k$ is indicated in Eq. (\ref{eq.gapnk}). }
 \label{tab.tab1}
 \begin{center}
\begin{tabular}{ccc}

\hline
\text{Dim(n+k)}&L(fm)&T(MeV)\\
3+1&1.198&164.7\\
2+2&1.379&143.1\\
1+3&1.484&133.0\\
0+4&1.557&126.7\\
\hline
\end{tabular}
\end{center}
\end{table}


\subsection{Stationary wave condition}

In the boundary of stationary wave condition, the dressed quark momentum is defined as
\begin{eqnarray}
 \vec{ p}_{SWC}^2&=&\frac{\pi^2}{L^2}\sum\nolimits_{i=1}^{3}n_i^2,\qquad n_i=+1,+2,+3...  \label{eq.psws}.
 \end{eqnarray}
For the case of SWC, there is no zero mode contribution. Here it should be noted that the zero mode is particularly important for the study of spontaneous symmetry breaking.

With this boundary condition, the dressed quark mass is given as
\begin{eqnarray}\label{eq.gap3}
M&=&m+48GM\int^ \infty_{\tau_{UV}}  d\tau e^{-\tau M^2}[T \times  \nonumber \\
& &\sum^\infty_{k=-\infty}e^{-\tau \omega_k^2}\prod \limits_{i=1}^3 \sum_{n_i  }  e^{-\tau   p_i^2}]  \nonumber  \\
&=&m+48GMT\int^ \infty_{\tau_{UV}} d\tau e^{-\tau M^2}(\frac{1} {2L})^3\times  \nonumber \\
&&\theta_2(0,e^{-4\pi^2\tau T^2}) [\theta_3(0,e^{-\tau \pi^2/L^2})-1]^3 ,
\end{eqnarray}
where $\theta_3(0,q)=1+2\sum\nolimits_{n=1}^\infty q^{n^2}$. Unlike Eq. (\ref{eq.gapnk}), Eq. (\ref{eq.gap3}) not only  shows $\theta_2$ function but also $\theta_3$ function.   Comparing Eq. (\ref{eq.gapnk}) and Eq. (\ref{eq.gap3}), we clearly see the difference in the dressed quark mass due to the use of the different APBC and SWC. And the strict mathematical equivalence of $T \sim 1/L$ is no longer established in this case. However, as our numerical calculations show below, using different boundary conditions only affects quantitatively rather than qualitatively the influence of finite volume effects on the chiral phase transition. For example, whether using SWC or the APBC, if the space size of the limited physical system is reduced, the chiral spontaneous symmetry breaking is prevented.

\begin{figure}
 { \includegraphics[width=0.8\columnwidth]{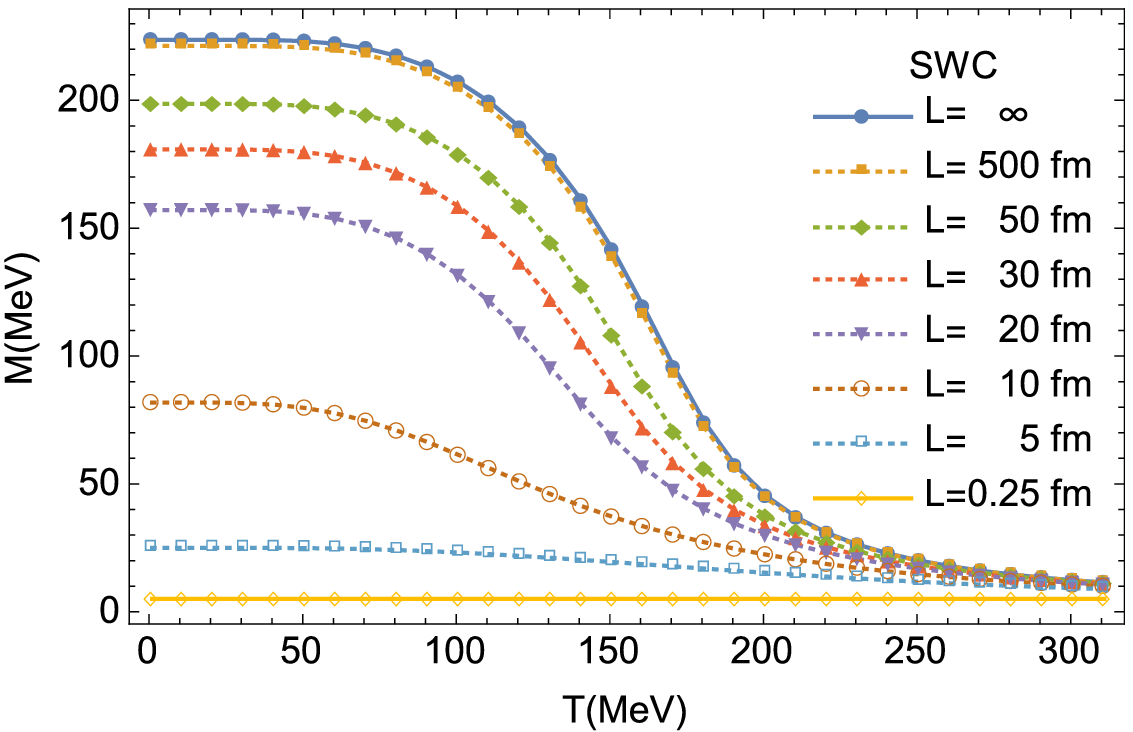}\\
 \includegraphics[width=0.8\columnwidth]{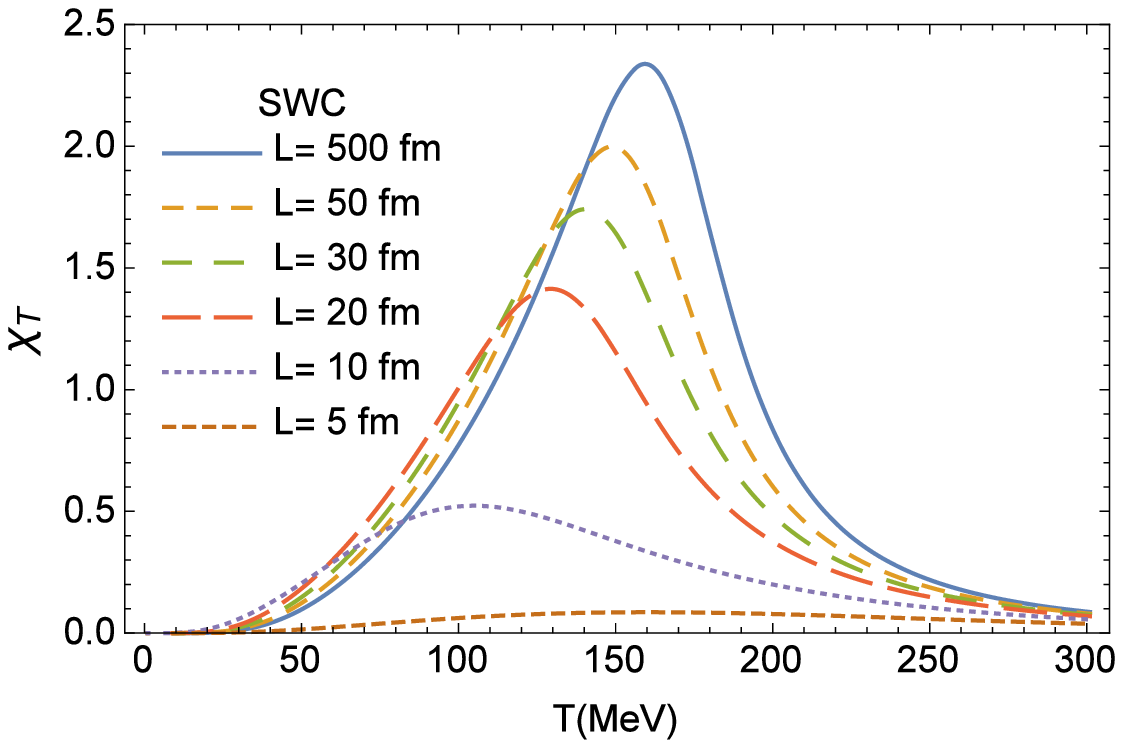}
 \caption{  Quark mass and  spatial susceptibility as function of temperature for different volume size with stationary wave condition.}
 \label{fig.mswc} }
\end{figure}

The dressed quark mass and the spatial susceptibility as function of temperature at different volume size is illustrated in Fig. \ref{fig.mswc}. Just as shown in Fig. \ref{fig.mswc}, for the stationary boundary condition, the dressed quark mass increases with increasing space size $L$. At the same time, the spatial susceptibility indicates the chiral phase transition caused by the small volume effects is a crossover. These phenomena are qualitatively similar to the case of APBC. From Fig. \ref{fig.mswc}, we can see that when $L$ is greater than $L^{SWC}_{max}=500$ fm the physical system of finite volume can be regarded as a thermodynamic system of infinite. Here we need to emphases that $500$ fm is far greater than the size of QGP produced at laboratory and the lattice QCD simulation space size. Fig. \ref{fig.mswc}  also tell us that when $L$ is less than $L^{SWC}_{min}=0.25$ fm, the dynamical chiral symmetry is effectively restored. What is particularly interesting is that $L^{SWC}_{min}=L^{APBC}_{min}$. Is this coincidence or something else worth further study? This is undoubtedly worth further study.
 
\section{summary}
\label{sec.summary}

Based on the two flavor NJL model with a proper time regularization, this work addresses finite volume effects with regard to the chiral phase transition at finite temperature.
we got the following interesting results: (1)  We are the first to use SWC in the world to study finite size effects in the chiral transition of QCD. Just as we shown in the present work  that SWC is quite different from APBC or PBC for the studying of finite size system. For example, the $L^{SWC}_{max}=500$ fm  given by SWC is far greater than that $L^{APBC}_{max}=5$ fm given by APBC. 
Given that the QGP system produced by the laboratory is very limited, the impact of the finite volume effects we derive from SWC far exceed the previous estimates made using the APBC. In addition, $L^{SWC}_{max}=500$ fm is also far greater than the maximum spatial simulation size of the Lattice QCD, which poses a challenge to the current attempt to use Lattice QCD to 
study the effect of the limited space size.
(2)  Similar to the temperature susceptibility, we first introduce the spatial susceptibility, and through the study of the spatial susceptibility, it was revealed for the first time that the chiral phase transition caused by the finite volume effects in the non-chiral limit is a smooth.
(3)  We first study the case that the three spatial directions are not all finite. It is found that the more discretized in the Euclidean space dimensions is, the faster the dressed quark mass declines.
(4)  For the first time, we use the two boundaries of APBC and SWC to  study  how small the physical systems ($L_{min}$ ) is the symmetry spontaneous breaking is no longer present? It is very interesting to find $L^{SWC}_{min}=L^{APBC}_{min}=0.25 fm$. What needs to emphasized here is that in the previous studies, people were mainly concerned with the influence of small volume effects on the CEP points in the QCD phase diagram., and never discussed the concept of spontaneous symmetry breaking is no longer valid under what circumstances. 
Finally, we need to emphasize that the above results are obtained in the framework of NJL model with the proper time regularization, and its effectiveness is worthy of further verification.

\section{Acknowledgement}
This work is supported in part by the National Natural Science Foundation of
China (under Grants  No.11690030, No.11475085,  No.11535005 ) and National Major state Basic Research and Development of China (2016YFE0129300)

\end{document}